# One shot profilometry using iterative two-step temporal phase-unwrapping

Guangliang Du[a], Minmin Wang[a], Canlin Zhou [a*], Shuchun Si[a], Hui Li[a], Zhenkun Lei[b], Yanjie Li[c]

[a] School of Physics, Shandong University, Jinan 250100, China

[b] Department of Engineering Mechanics, Dalian University of Technology, Dalian 116024, China

[c] School of Civil Engineering and Architecture, University of Jinan, Jinan, 250022, China

[*]Corresponding author: Tel: +8613256153609; E-mail address: canlinzhou@sdu.edu.cn

## Abstract

This paper reviews two techniques that have been recently published for 3D profilometry and proposes one shot profilometry using iterative two-step temporal phase-unwrapping by combining the composite fringe projection and the iterative two-step temporal phase unwrapping algorithm. In temporal phase unwrapping, many images with different frequency fringe pattern are needed to project which would take much time. In order to solve this problem, Ochoa proposed a phase unwrapping algorithm based on phase partitions using a composite fringe, which only needs projecting one composite fringe pattern with four kinds of frequency information to complete the process of 3D profilometry. However, we found that the fringe order determined through the construction of phase partitions tended to be imprecise. Recently, we proposed an iterative two-step temporal phase unwrapping algorithm, which can achieve high sensitivity and high precision shape measurement. But it needs multiple frames of fringe images which would take much time. In order to take into account both the speed and accuracy of 3D shape measurement, we get a new, and more accurate unwrapping method based on composite fringe pattern by combining these two techniques. This method not only retains the speed advantage of Ochoa's algorithm, but also greatly improves its measurement accuracy. Finally, the experimental evaluation is conducted to prove the validity of the proposed method, and the experimental results show that this method is feasible.

phase-unwrapping

## 1. Introduction

Fringe-projection profilometry(PFP) is a well-developed technique and powerful tool in current three-dimensional shape measurement for its non-contact nature,full-field measurement capability, high profile sampling desnsity and low environment vulnerability, which has been an extensively studied research area due to the diversity of potential application[1-4]. The basic operation of the process is shown below, first of all, sinusoidal fringes are projected onto an object surface, images of the fringe patterns deformed by the object surface are captured by a camera, then a phase map is calculated from the pixel in the images. Because the retrieved phase distribution corresponding to the object height is wrapped in the range $(-\pi, \pi)$, phase unwrapping has become one of the key steps in grating projection 3D shape measurement[5-7]. In recent decades, a variety of phase unwrapping algorithms have been proposed, they can be divided into two categories: spatial phase unwrapping method and temporal phase unwrapping method. Spatial phase unwrapping is a process of integral accumulation, once the error appears, it will spread to the point around, so noise, shadow and discontinuous points in the actual measurement will affect the phase unwrapping quality[8-9]. Temporal phase unwrapping method can solve this problem. Temporal phase unwrapping method is proposed by Saldner et al.[10], which is realized by projecting a series of different frequency grating to the surface of the measured object, capturing the fringe sequence modulated by the object surface, then unwrapping the phase along the time series for each pixel independently, so it can avoid the phase unwrapping error propagation. However, it is unable to meet the requirements of real-time dynamic 3-D shape measurement. How to realize the phase unwrapping algorithm in time domain with fewer projection fringes is concerned by everyone. Liu[11] proposed the dual-frequency method, which combines a unit-frequency fringe pattern with a high-frequency. Liu[12] proposed tri-Frequency Heterodyne Method. Gai proposed the amplitude modulation method, where the unit frequency is obtained from amplitude modulation of the fringes [13]. Manuel Servin proposed a 2-step temporal phase unwrapping algorithm [14], which only needs the 2 extreme phase-maps to achieve exactly the same results as standard temporal unwrapping method. Recently, we proposed an iterative two-step temporal phase unwrapping algorithm[15] to achieve high sensitivity and high precision shape measurement. In order to further improve the measuring speed, Wei[16] proposed a single-shot measurement method, which requires an unwrapping process for both the high and low frequency phases. Ochoa[17] proposed a phase unwrapping algorithm based on phase partitions using a composite fringe, which only needs projecting one composite fringe pattern with four kinds of frequency information to complete the process of unwrapping. However, according to our own experiments with Ochoa's method, we found that the fringe order determined through the construction of phase partitions tended to be imprecise. In order to take into account both the speed and accuracy of 3D shape measurement, we get a new, and more accurate unwrapping method based on composite fringe pattern by combining the composite fringe projection in [17] and the iterative two-step temporal phase unwrapping algorithm[15]. This method not only retains the speed advantage of Ochoa's algorithm, but also greatly improves its measurement accuracy. The capability of the presented method is demonstrated by both theoretical analysis and experiments.

The paper is organized as follows. Section 2 reviews these two methods and combines these two techniques into a new and more accurate 3D profilometry. Section 3 presents the experimental results. Section 4 summarizes this paper.

## 2. Theory

### 2.1 Phase partitions unwrapping algorithm using a composite fringe

Ochoa et al. proposed a phase partitions unwrapping algorithm using a composite fringe pattern, the composite pattern to be projected is described by the equation below,

$$I(x,y) = \frac{G}{8}\{4 + \cos(2\pi f_1 x) + \cos(2\pi f_2 x) + \cos(2\pi f_2 y) + \cos(2\pi (f_2+1)x + 2\pi f_2 y)\} \quad (1)$$

where $f_1$ and $f_2$ are medium and high carrier frequencies $(2f_1 < f_2)$, $G$ is a constant that represents the maximum gray level range, $(x,y)$ are the normalized pixel coordinates, and $I(x,y)$ is the image with its gray levels in the range $[0, G]$. The four carrier terms are given by,

$$\begin{aligned} C_{x1}(x,y) &= 2\pi f_1 x; \\ C_x(x,y) &= 2\pi f_2 x; \\ C_y(x,y) &= 2\pi f_2 y; \\ C_{xy}(x,y) &= 2\pi (f_2+1)x + 2\pi f_2 y. \end{aligned} \quad (2)$$

Thus the following relation should hold[18],

$$C_{xy}(x,y) - C_x(x,y) - C_y(x,y) = 2\pi x \quad (3)$$

which is a unit frequency vertical fringe.

The intensity profile that we will obtain after projecting Eq.(1) onto the object's surface will be given by

$$I = a + b[\cos(C_{x1} + \varphi^{x1}) + \cos(C_x + \varphi^x) + \cos(C_y + \varphi^y) + \cos(C_{xy} + \varphi^{xy})] \quad (4)$$

where $a$ and $b$ are background and amplitude terms that depend on the object's reflectivity, respectively, and $\varphi^{x1}$, $\varphi^x$, $\varphi^y$ and $\varphi^{xy}$ are the phase functions related to the surface height. Ochoa used Fourier analysis to demodulate the composite fringe patterns and obtain the wrapped phase $W[C_{x1} + \varphi^{x1}]$, $W[C_x + \varphi^x]$, $W[C_y + \varphi^y]$ and $W[C_{xy} + \varphi^{xy}]$, where $W[.]$ is the wrapping phase operator. According to Eq.(8)-Eq.(9) in [17], we can obtain the vertical phase $2\pi x + \varphi^{eq}(x,y)$ only with one period, , where $\varphi^{eq} = \varphi^{xy} - \varphi^x - \varphi^y$ represents the equivalent phase (unwrapped) of the phase differences. In order to unwrap the high-frequency phase $W[C_x + \varphi^x]$, Ochoa proposed a phase partitions unwrapping algorithm. First, determine $n$ and $m$ by the following relationship,

$$\begin{aligned} \left(\frac{n(x,y)-1}{f_1}\right)2\pi &\leq [2\pi x + \varphi^{eq}(x,y)] < \frac{n(x,y)}{f_1}2\pi \\ \left(\frac{m(x,y)-1}{f_2}\right)f_1 2\pi &\leq W[C_{x1} + \varphi^{x1}(x,y)] < \frac{m(x,y)}{f_2}f_1 2\pi \end{aligned} \quad (5)$$

$n(x,y)$, $m(x,y)$ are integer values and $n \in [0, f_1-1]$, $m \in [0, (f_2/f_1)-1]$, then the

fringe order of $W[C_x + \varphi^x]$ can be determined as

$$N(x,y) = n(x,y)f_1 + m(x,y) \qquad (6)$$

which let us calculate the absolute the high-frequency phase term through the following equation

$$\varphi^x(x,y) = W[C_x + \varphi^x(x,y)] + 2\pi N(x,y) - 2\pi fx \qquad (7)$$

As noted in reference[17], some erroneous fringe order can be generated because of the Fourier transform and filtering operation. So the unwrapped phase should be corrected by the algorithm in [19]. Ochoa's method can calculate the unwrapped phase and the profile information of discontinuous, isolated and complex objects, when only one shot is used through Fourier techniques, which has the advantage of speed and suits for fast 3D surface measurement. However, according to our own experiments with Ochoa's method, we found that the fringe order determined through the construction of phase partitions tended to be imprecise. Even with the correction method in [19][20], the imprecise fringe order can not be completely eliminated, leading to many errors in the final unwrapped phase.

### 2.2 Iterative two-step temporal phase unwrapping algorithm

Recently, we proposed an iterative two-step temporal phase unwrapping algorithm[15], to achieve high sensitivity and high precision shape measurement. Assuming that the intensity mathematical formula for three fringe patterns with different phase modulation sensitivities is as follows,

$$\begin{aligned}
&I1(x,y) = a(x,y) + b(x,y)\cos[\varphi(x,y)], \varphi(x,y) \in (-\pi, \pi), \\
&I2(x,y) = a(x,y) + b(x,y)\cos[g_1\varphi(x,y)], (g_1 > 1), g_1 \in R, \\
&I3(x,y) = a(x,y) + b(x,y)\cos[g_1*g_2\varphi(x,y)], (g_2 > 1), g_2 \in R.
\end{aligned} \qquad (8)$$

where $\varphi(x,y)$ is a 1λ sensitive phase (λ is wavelength) and $G\varphi(x,y)$ is G-times more sensitive, $G = g_1 * g_2$. We can use the phase demodulation algorithm to obtain the 3 demodulated wrapped phase-maps as,

$$\begin{aligned}
&\varphi 1(x,y) = W[\varphi(x,y)], \varphi(x,y) \in (-\pi, \pi) \\
&\varphi 2_w(x,y) = W[g_1\varphi(x,y)], (g_1 > 1), g_1 \in R, \\
&\varphi 3_w(x,y) = W[g_1 g_2 \varphi(x,y)], (g_2 > 1), g_2 \in R.
\end{aligned} \qquad (9)$$

The first demodulation $\varphi 1(x,y)$ is not wrapped because it is less than 1λ. So, we have, $\varphi 1(x,y) = \varphi(x,y)$. $\varphi 2_w(x,y)$ and $\varphi 3_w(x,y)$ are the wrapped phase, and we can unwrap them by,

$$\begin{aligned}
&\varphi 2(x,y) = g_1 \varphi 1(x,y) + W[\varphi 2_w(x,y) - g_1 \varphi 1(x,y)] \\
&\varphi 3(x,y) = g_2 \varphi 2(x,y) + W[\varphi 3_w(x,y) - g_2 \varphi 2(x,y)]
\end{aligned} \qquad (10)$$

Due to the use of high sensitivity fringe, the iterative two-step temporal phase unwrapping algorithm can increase the signal-to-noise power-ratio, achieve higher sensitivity and more accurate measurement, and suit for isolated objects. But it needs multiple frames of fringe images which would take much time.

## 2.3 One shot profilometry using iterative two-step temporal phase-unwrapping

With the widely used of 3-D surface measurement, the speed and accuracy of the measurement should be higher and higher. The phase partitions unwrapping algorithm proposed by Ochoa, can improve the measurement speed, but the results are imprecise; and the iterative two-step temporal phase unwrapping algorithm can achieve higher sensitivity and more accurate measurement, but it needs multiple frames of fringe images. We combine these two algorithms and proposed a more effective 3-D surface measurement method. This method not only retains the speed advantage of Ochoa's algorithm, but also has the accuracy advantage of the iterative two-step temporal phase unwrapping algorithm. The main stages of our algorithm are summarized as follows：

(1) project Eq.(1) onto the object's surface and capture the image by a CCD camera, as shown in Fig.1-A;

(2) obtain the wrapped phase $W[C_{x1}+\varphi^{x1}]$, $W[C_x+\varphi^x]$, $W[C_y+\varphi^y]$ and $W[C_{xy}+\varphi^{xy}]$ by Fourier transform, filtering operation ,inverse Fourier transform, as shown in Fig.1-B,C;

(3) obtain the unit-frequency phase $2\pi x+\varphi^{eq}(x,y)$ according to Eq.(8)-Eq.(9) in ref.[17], as shown in Fig.1-D;

(4) unwrap the high-frequency phase $W[C_x+\varphi^x]$ with the iterative two-step temporal phase unwrapping algorithm Eq.(10), obtain the continuous phase $C_x+\varphi^x$, as shown in Fig.1-E,F;

(5) transform the absolute phase to height information after the system is calibrated.

The specific procedure is shown in Fig. 1. As you might see, the main stages of proposed method are same as that of Ochoa's algorithm except that the partitions unwrapping algorithm is replaced with the iterative two-step temporal phase unwrapping algorithm.

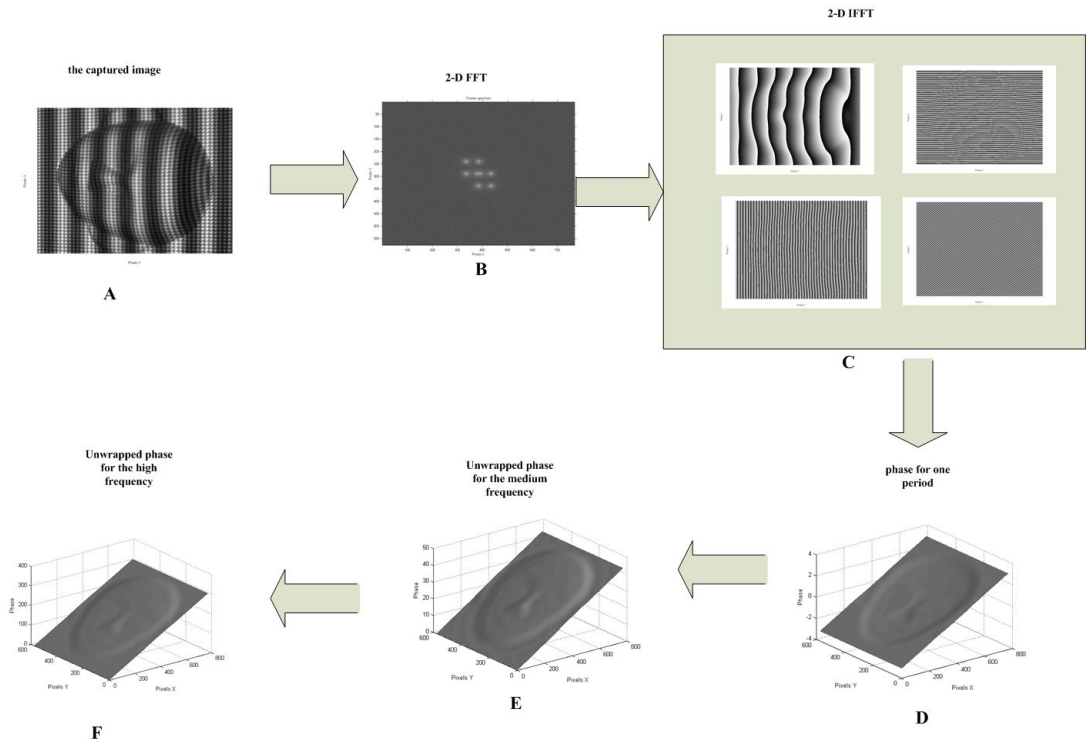

Fig.1 The specific procedure

The following experiment is used to verify the proposed algorithm.

## 3. Experiments

In this section, for evaluating the real performance of our method, we test our method on a series of experiments. Below, we will describe these experiments and practical suggestions for the above procedure.

We develop a 3D shape measurement system, which consists of a DLP projector (Optoma EX762) driven by a computer and a CCD camera ( DH-SV401FM).The camera is attached with a 25mm focal length lens(Model: ComputarFA M2514-MP2). Fig.2 shows the optical path of phase measuring profilometry, where P is the projection center of the projector, C is the camera imaging center, and D is an arbitrary point on the tested object. The surface measurement software is programmed by Matlab with I5-4570 CPU @ 3.20 GHz.

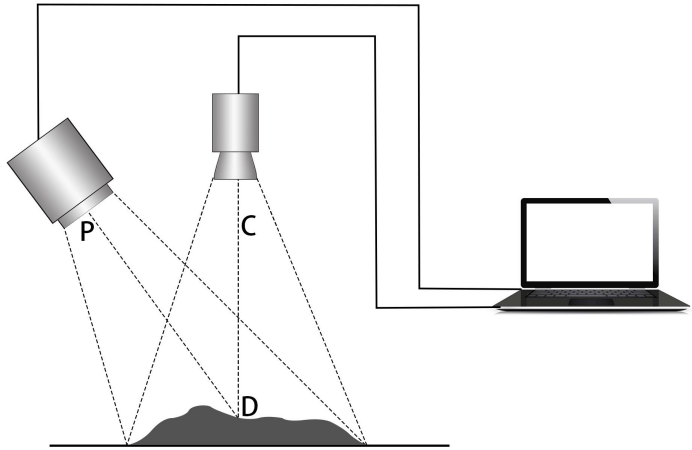

Fig. 2 The optical path of phase measuring profilometry

Firstly, an experiment is provided to demonstrate the feasibility of the proposed algorithm. The tested object is a face model and an isolated cup. As said in[17], there will be loss of information due to the Fourier process, so we start with phase shift method to prove our approach. In this paper, we use the four-step phase shift method. The four-step phase shift fringes are projected onto the object surface, the captured images are shown in Fig.3, and the image is 592 pixels wide by 496 pixels high. Fig.3(a) is the vertical fringe pattern with 7 periods, Fig.3(b), Fig.3(c) are the vertical, horizontal fringe pattern with 49 periods respectively, Fig.3(d) is tilted fringe pattern whose vertical and horizontal periods are 49,50 respectively.

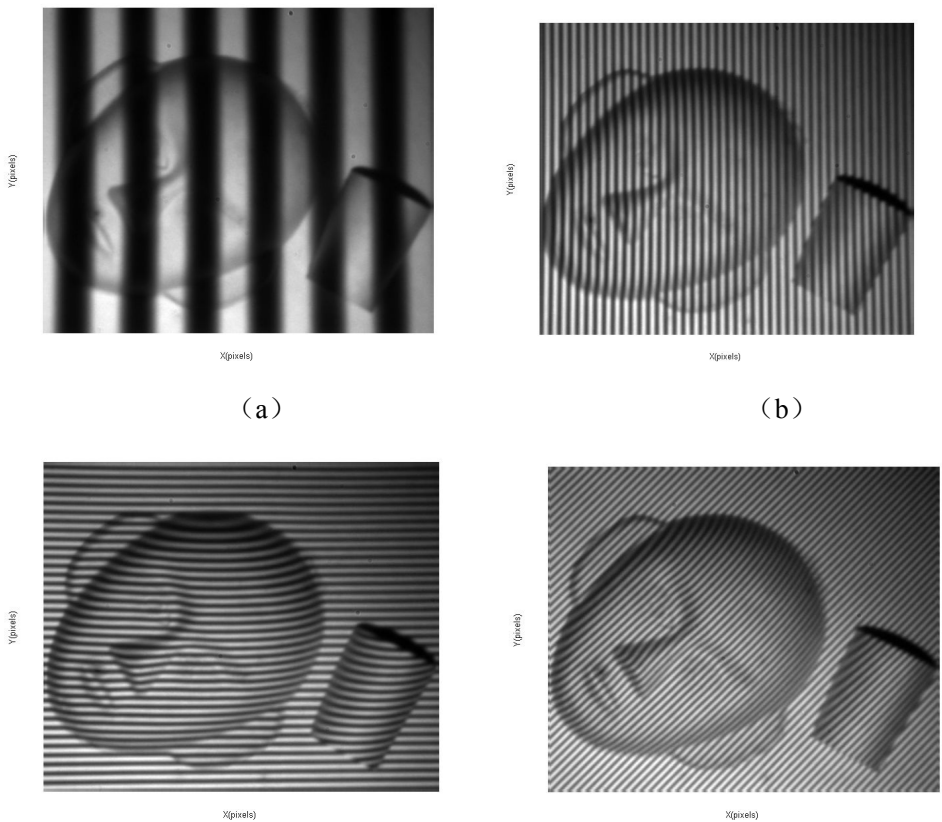

（a）　　　　　　　　　　　　　　（b）

（c） （d）

Fig.3 The captured images

Fig.4 is the wrapped phase obtained by four-step phase shift method. Fig.4(a), Fig.4(b), Fig.4(c), Fig.4(d) are the wrapped phases of Fig.3(a), Fig.3(b), Fig.3(c), Fig.3(d), respectively.

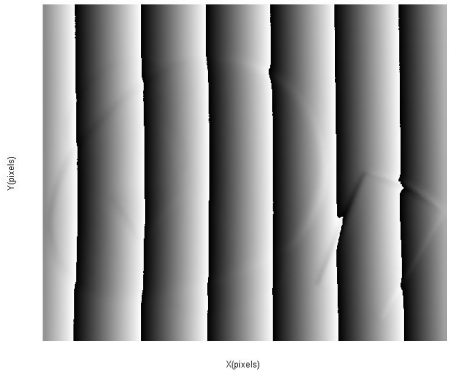
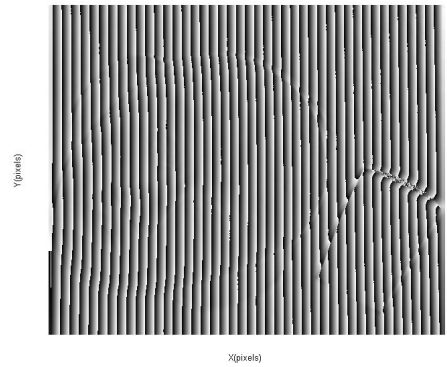

（a） （b）

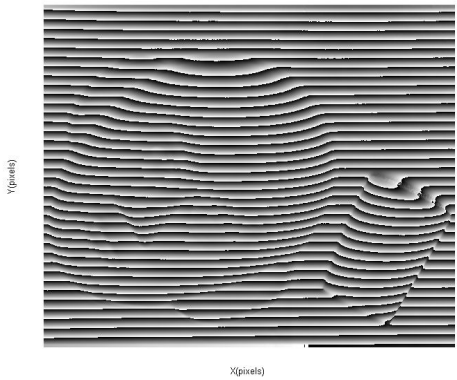
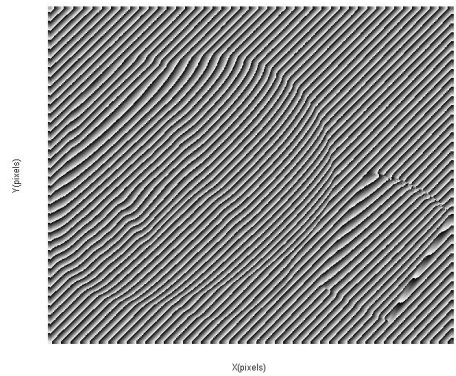

（c） （d）

Fig.4 The wrapped phase obtained by four-step phase shift method

The vertical phase with one period $2\pi x + \varphi^{eq}(x, y)$ is shown in Fig.5.

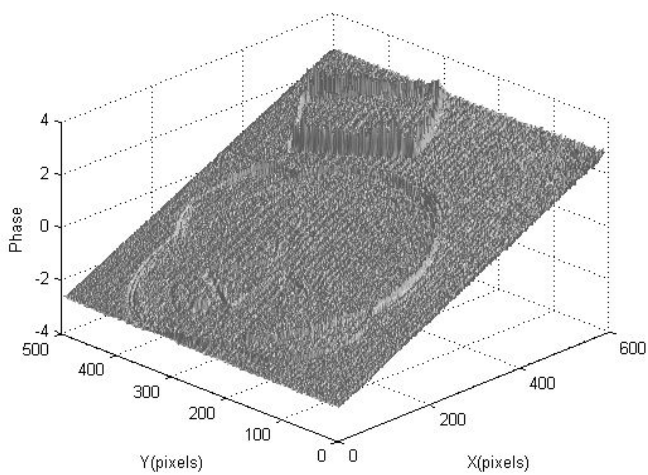

Fig.5 The vertical phase with one period

First, dealing with the captured images in Fig.3 by the Ochoa's method using Eq.(5)-(7), the result is shown in Fig.6.

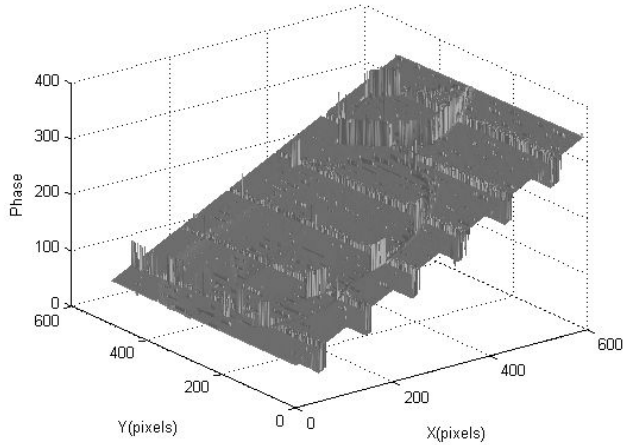

Fig.6 The unwrapped phase by Ochoa's algorithm

Then, we correct the fringe order using the method disussed in references[19] [20].The results are shown in Fig.7. Fig.7(a) and Fig.7(b) are the corresponding results after two correction methods are used to modify the mistaken fringe order. As one might see, there are still obvious errors in retrieved phase map.

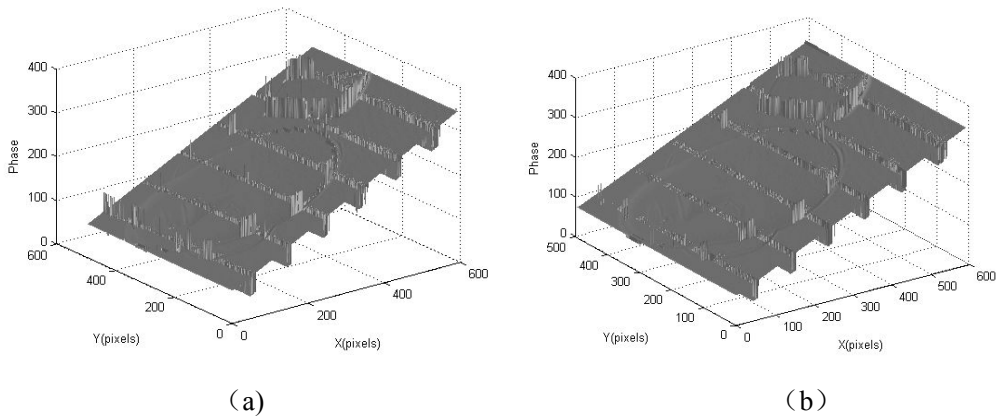

（a）　　　　　　　　　　　　　　（b）
Fig.7 The unwrapped phase after correction

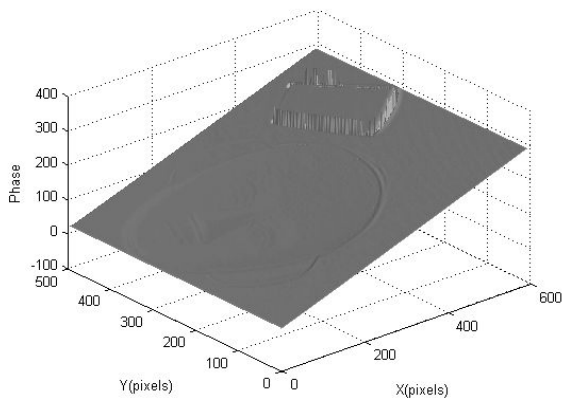

Fig.8 The unwrapped phase obtain by proposed method

In contrast, dealing with the captured images in Fig.3 by the proposed method, we get the result shown in Fig.8. It is obvious that the proposed method can obtain the precise unwrapped phase while Ochoa's algorithm can introduce large error.

For a more complete verification to the proposed method, we do another experiment on a plastic board with a big hole. To validate that the method is applied to objects with hole, we use a black background in the hole area, as shown in Fig.9. The experiment procedure is similar. But we use Fourier method and composite fringe projection.

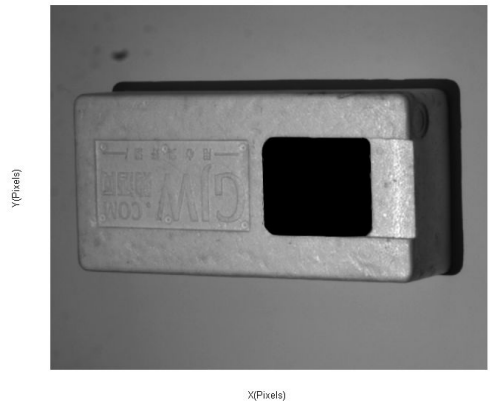

Fig.9 The tested object

Project the composite fringe produced according to Eq.(1) onto the object's surface, the deformed fringe pattern is captured by a CCD camera ( $f_1 = 7, f_2 = 49$ ), as shown in Fig.10. The captured image is 592 pixels wide by 496 pixels high.

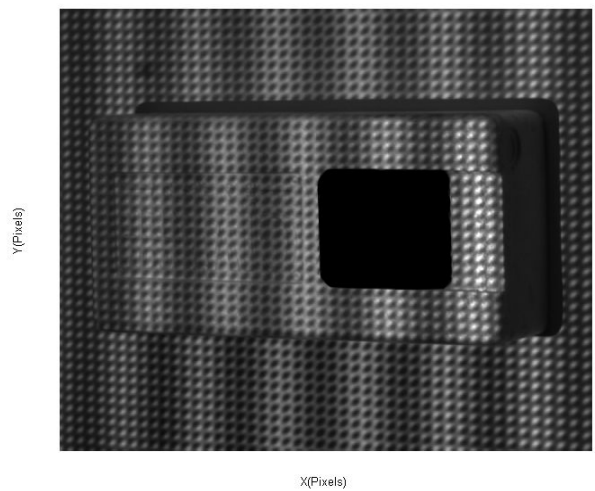

Fig.10 The captured image

Then the image can be analyzed to get the 3D surface shape, as described in section 2.2 and 2.3. The spectrum of the captured image by 2D Fourier transform is shown in Fig.11.

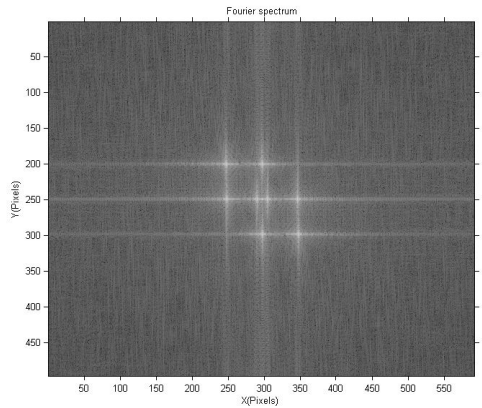

Fig.11 The spectrum of the captured image

Fig.12 is the wrapped phase obtained by filtering operation. Fig.12(a) is the vertical wrapped phase with 7 periods, Fig.12(b), Fig.12(c) are the vertical, horizontal wrapped phase with 49 periods respectively, Fig.12(d) is tilted wrapped phase whose vertical and horizontal periods are 49,50 respectively.

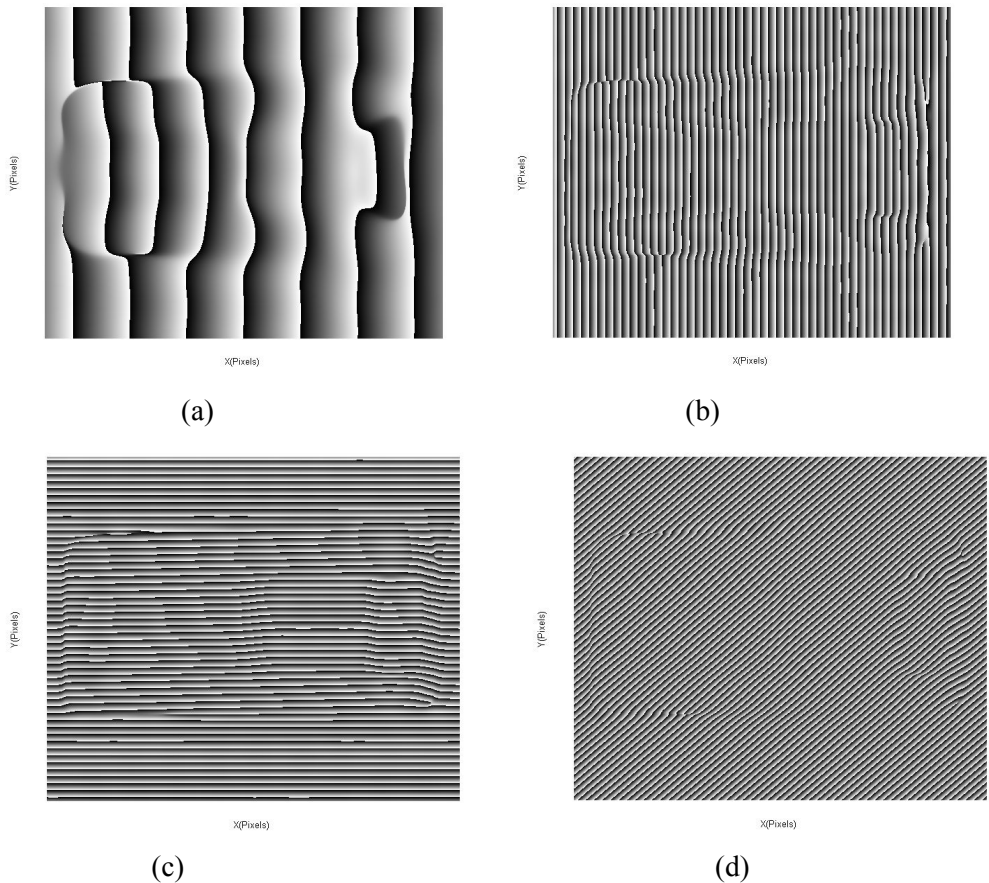

(a)          (b)

(c)          (d)

Fig.12 The wrapped phase obtained by filtering operation

First, we reconstruct the 3D shape from the wrapped phase shown in fig.12 using phase partitions method, The result is shown in Fig.13.

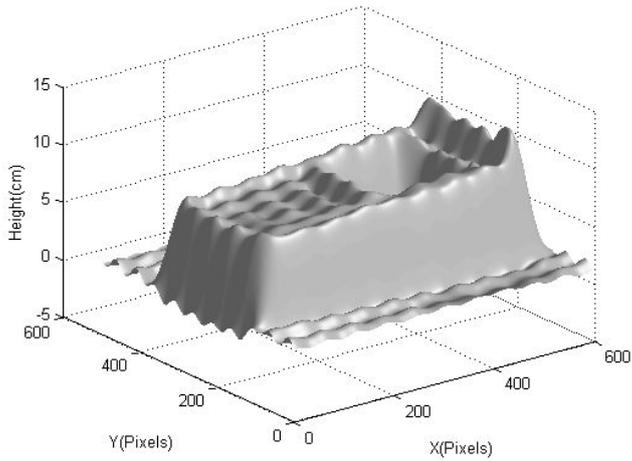

Fig.13 The 3D reconstructions using Ochoa's algorithm

Then，the high frequency phase is independently unwrapped from the wrapped phase shown in fig.12 by the proposed algorithm,The result is illustrated in Fig.14. As shown in Fig.13 and 14, the proposed method can restore 3D surface shape well, while Ochoa's method leads to many errors in the final result.

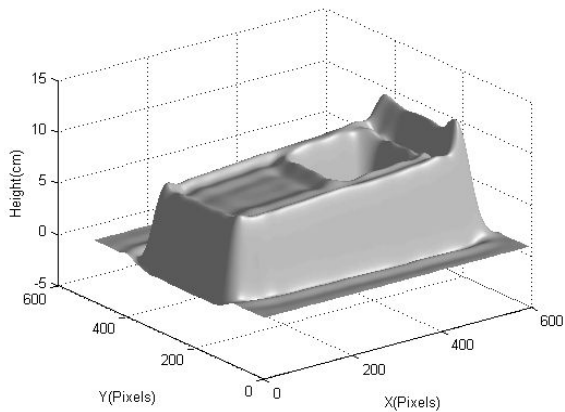

Fig.14 The 3D reconstructions using the proposed algorithm

To have a better comparison, the results of the 200th row from Fig.13 and 14 are provided in Fig.15, Fig.15(a) is the result processed by Ochoa's method. Fig. 15(b) is the result processed by the proposed method.It can be seen from Fig.15 that the proposed method provides the higher success rate as expected.

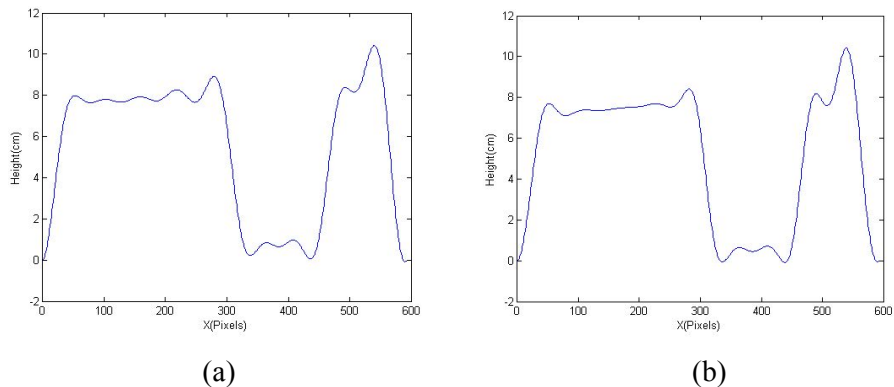

(a)                               (b)

Fig.15 The results of the 200th rows of the two methods

It is well known that some information will be lost due to the spectral leakage, the selection of spectral filtering window and other reasons in the process of Fourier transform. And some erroneous values will be generated especially at the borders. Phase shift method can solve this problem by increasing the number of projected images.

## 4. Conclusion

In this paper, we propose an one shot profilometry using iterative two-step temporal phase-unwrapping by combining the composite fringe projection and the iterative two-step temporal phase unwrapping algorithm. The proposed method can calculate the unwrapped phase and restore 3D surface shape of isolated and complex objects, meanwhile it needs only one shot. And it not only retains the speed advantage of Ochoa's algorithm, but also greatly improves its measurement accuracy.

**Acknowledgment**
This work was supported by the National Natural Science Foundation of China (Grant nos. 11302082 and 11472070). The support is gratefully acknowledged. We would like to express our gratitude to Noé Alcalá Ochoa for invaluable help and useful discuss.